\documentclass[12pt,preprint]{aastex}

\shorttitle{Estimating Column Density in Molecular Clouds}
\shortauthors{Schnee, Bethell \& Goodman}
\begin{document}

\title{Estimating Column Density in Molecular Clouds with FIR and Sub-mm 
	Emission Maps}

\author{S. Schnee$^1$, T. Bethell$^2$ \& A. Goodman$^1$}
\affil{$^1$Harvard-Smithsonian Center for Astrophysics, 60 Garden
       Street, Cambridge, MA 02138 \\ 
       $^2$Department of Astronomy, University of Wisconsin, Madison, 
       5534 Sterling Hall, 475 North Charter Street, Madison, WI 53706}
\email{sschnee@cfa.harvard.edu}

\begin{abstract}
	We have used a numerical simulation of a turbulent cloud to
synthesize maps of the thermal emission from dust at a variety of
far-IR and sub-mm wavelengths.  The average column density and
external radiation field in the simulation is well matched to clouds
such as Perseus and Ophiuchus.  We use pairs of single-wavelength
emission maps to derive the dust color temperature and column density,
and we compare the derived column densities with the true column
density.  We demonstrate that longer wavelength emission maps yield
less biased estimates of column density than maps made towards the
peak of the dust emission spectrum.  We compare the scatter in the
derived column density with the observed scatter in Perseus and
Ophiuchus.  We find that while in Perseus all of the observed scatter
in the emission-derived versus the extinction-derived column density
can be attributed to the flawed assumption of isothermal dust along
each line of sight, in Ophiuchus there is additional scatter above
what can be explained by the isothermal assumption.  Our results imply
that variations in dust emission properties within a molecular cloud
are not necessarily a major source of uncertainty in column density
measurements.
\end{abstract}


\section{Introduction}
\label{INTRO}
	In the absence of direct meausurements of extinction, the
column density of dust in molecular clouds is often traced by dust
emission.  In many cases the data are maps at two wavelengths in the
FIR or sub-mm, and the column density is derived by assuming that dust
along each line of sight is isothermal
\citep{Wood94,Schlegel98,Schnee05}.  However, each line of sight
through a molecular cloud is likely to intersect regions with various
amounts of shielding, and therefore a range of temperatures, so the
isothermal assumption may lead to significant errors in the
calculation of column density.
	
	In this paper we use a simulation of an externally heated
turbulent molecular cloud to produce maps of column density and
thermal emission at various FIR and sub-mm wavelengths.  We discuss
the bias in the emission-derived column density as a function of
extinction and wavelength, and we compare the scatter in the
emission-based column density estimate with that observed in
\citet{Schnee05} in the Perseus and Ophiuchus molecular clouds.  We
find that assuming isothermality can account for all of the observed
point by point scatter in comparisons of extinction and emission-based
column density maps of Perseus without the need to invoke variations
in the dust properties within a molecular cloud or variations in the
emission from stochastically heated grains.  In comparison, the
scatter in Ophiuchus is significantly larger than can be accounted for
by the error introduced by the isothermal dust assumption alone.

\section{The Model Cloud}                                                   
\label{MODEL}      
	The method of determining the dust distribution, radiative
transfer and dust heating, only briefly described in this Section, are
described in more detail in \citet{Bethell04}.

\subsection{Density Structure}      
	The density structure is created by a three-dimensional
simulation of driven MHD turbulence on a cubical grid with $128$ cells
on a side \citep{Heitsch01}.  Although the original ensemble of
simulations was created with widely varying physical parameters (Mach
number, plasma beta etc.), the radiative transfer through the
resulting structures is largely insensitive to these differences.  To
simulate Perseus and Ophiucus we use one such model and scale the
density and size of our model cloud so that the column density map has
an average $A_V \sim 2$, which is the median $A_V$ in our Perseus and
Ophiuchus extinction maps.  This choice of model is meant to be
illustrative of the broad effects of including `realistic' clumpiness
in radiative transfer calculations of molecular clouds.  When compared
to a uniform density distribution of similar mass the inclusion of
clumpiness tends to expose more of the \textit{volume} to the
interstellar radiation field (ISRF) while on average the \textit{mass}
is both less illuminated and sees bluer light.  In the clumpy case
there is also little correlation between the intracloud radiation
field at a point and its proximity to the cloud edge. Instead, there
is a large point-to-point scatter, arising from the details of the
clumpiness \citep{Bethell04}.

\subsection{Radiative transfer}    
	A Reverse Monte Carlo radiative transfer code
\citep{Bethell04} is used to calculate the penetration of an external
isotropic and homogeneous ISRF, here the ISRF given by
\citet{Mathis83}.  The reverse Monte Carlo technique enables the user
to control the Monte Carlo photon noise in regions which are well
shielded, in particular the clumps where most of the mass resides.
The re-emission by dust of sub-mm radiation is assumed to be optically
thin throughout our cloud.

\subsection{Dust properties}                                                 
	To calculate self consistently the attenuation of the ISRF by
dust, the subsequent equilibrium dust temperature and the FIR emission
we use astronomical silicate and graphite grain opacities
\citep{Draine84}\footnote{For tabulated optical properties visit the
website www.astro.princeton.edu/$\sim$draine.} and a power-law grain
size distribution \citep{MRN1977}.  These are chosen to reproduce an
extinction curve with $R_{V}=A_{V}/E(B-V)=3.1$.  No attempt is made to
include very small grains which are heated transiently to high
temperatures, nor do we include polycyclic aromatic hydrocarbon
emission.  Both these populations contribute significantly to
wavelengths shorter than 100 \micron.

	At far infrared wavelengths ($\lambda > 100 \mu m$) the dust
ensemble emission is described by a blackbody spectrum modified by a
power law ($S_\lambda \propto B_\lambda(T_d)\lambda^{-\beta})$ with an
emissivity spectral index of $\beta=2$.  With the dust ensemble fully
prescribed we can calculate the equilibrium grain temperatures of each
grain type and grain size at each location.  Not only do dust
temperatures vary along each line of sight, they also vary amongst
grain types and sizes at any point within the cloud.

\section{Derived Color Temperature and Column Density}
\label{DERIVED}
	The method used to determine the dust temperature and column
density from two emission maps is similar to that used in
\citet{Schnee05}.  The dust temperature ($T_d$) is determined by the
ratio of two flux densities and the column density can be derived from
either flux and the derived color temperature of the dust.  The pairs
of fluxes that we study in this paper are 60 and 100 \micron\ (the
longest wavelength IRAS bands), 100 and 240 \micron\ (those used in
\citet{Schlegel98}, hereafter SFD98), 100 and 850 \micron\ and 450 and
850 \micron\ (wavelengths observed by SCUBA on the JCMT
\citep{Holland99}).

	To determine the dust color temperature, we use the equation:
\begin{equation}
\label{TEMPEQN}
R=\left(\frac{\lambda_1}{\lambda_2}\right)^{-(5+\beta)}
  \left(\frac{\exp[hc/\lambda_2 k T_d]-1}{\exp[hc/\lambda_1 k T_d]-1}\right)
\end{equation}
where R is the ratio of the fluxes and $\beta$ is the emissivity
spectral index of the dust.  We calculate the dust temperature with
the assumption that $\beta = 2$, which is true of the dust in our
model.  Note that in using the ratio of two flux maps to get the color
temperature, we are assuming that a single temperature, $T_d$, can be
used to describe all the dust along each line of sight.

	The optical depth is determined by the ratio of the flux to
the Planck function:
\begin{equation}
\label{TAUEQN}
\tau_\lambda=\frac{F_\lambda}{B_\lambda(T_d)}
\end{equation}
with the assumption that the dust emission is optically thin.

	The optical depth can then be converted to V-band extinction
using:
\begin{equation}
\label{AVEQN}
A_V=X \tau_\lambda
\end{equation}
where $X$ is a parameter relating the thermal emission properties of
dust to its optical absorption qualities.  The extinction, $A_V$, is
derived from the column density of dust in our model, using the
conversions $N_H/E(B-V) = 5.8 \times 10^{21}$ cm$^{-2}$ and
$A_V/E(B-V) = 3.1$.  We use the IDL procedure AMOEBA to determine the
value of $X$ that minimizes the difference between the
emission-derived column density and the true column density in the
simulation.

\section{Results}
\subsection{Dependence on Wavelength}
\label{WAVELENGTH}
\begin{figure}
\epsscale{1.0}
\plotone{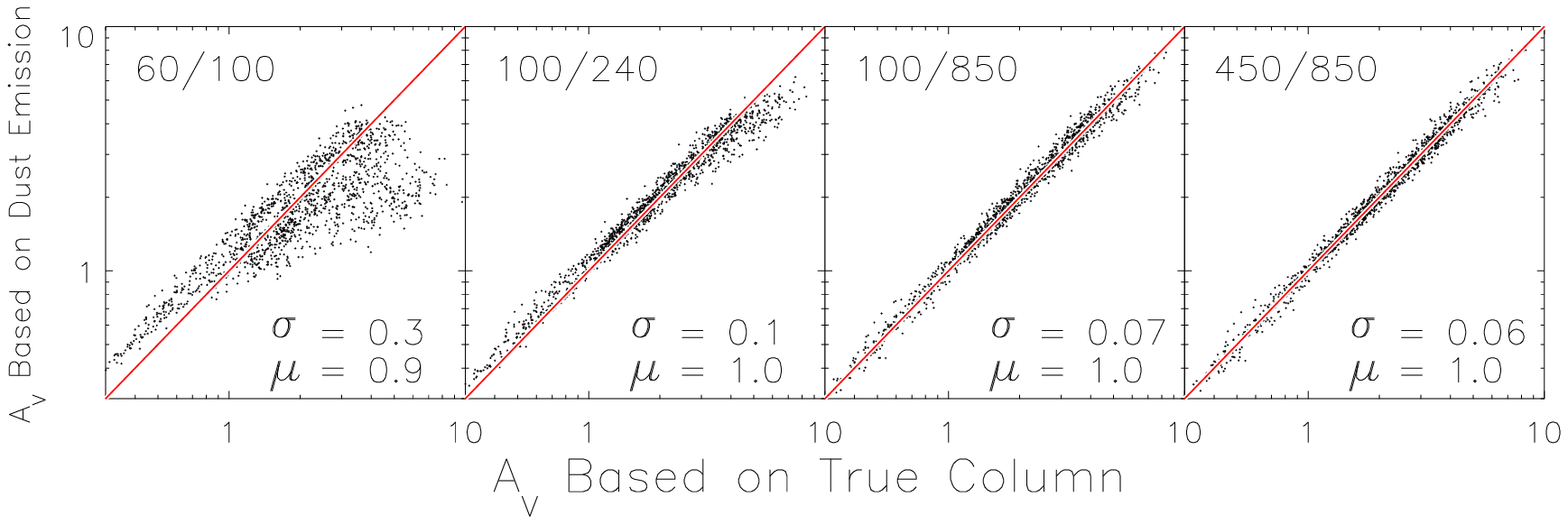}
\caption{The column density derived from two emission maps plotted
         against the true column density.  The wavelengths used are,
         from left to right: 60 \& 100 \micron, 100 \& 240 \micron, 
	 100 \& 850 \micron\ and 450 \& 850 \micron.  The values of $X$ 
	 that convert from the FIR optical depth to visual extinction
         are $\tau_{100} = 775$, $\tau_{100} = 231$, $\tau_{100} = 69$, 
         and $\tau_{450} = 681$.  The values of $\mu$ and $\sigma$ refer
         to the center and width of the gaussian fit to the distribution
         of the ratio ($A_{V_{emission}}/A_{V_{true}}$).
	 \label{FIG1}}
\end{figure}

	The column density of dust derived from pairs of thermal
emission maps depends strongly on the two wavelengths considered.
Figure \ref{FIG1} shows that the 60 and 100 \micron\ derived column
density is a linear tracer of the actual column density for $A_V < 1$
mag, but systematically underestimates the true column density for
$A_V > 1$ mag.  The same trend is seen in the column density derived
from the 100 and 240 \micron\ emission maps, though the cutoff
extinction above which the column densities are underestimated is $A_V
\sim$4 mag.  The column density derived from the 100 and 850 \micron\
maps traces the true column density all the way to 10 mag $A_V$, as
does the column density derived from the 450 and 850 \micron\ emission
maps.  The errors introduced by the assumption of isothermal dust
along each line of sight for each pair of maps is shown in Figure
\ref{FIG1}.

\begin{figure}
\epsscale{0.75}
\plotone{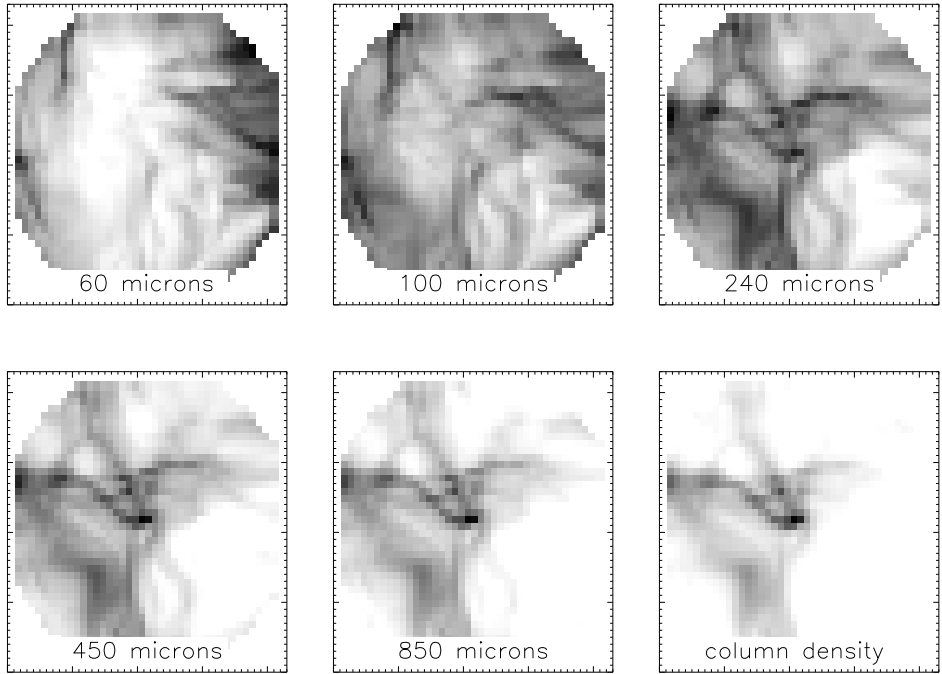}
\caption{The flux maps at the five wavelengths considered in the paper,
         plotted along with the true column density.  All of the 
         maps are scaled to have the same dynamic range with their maxima 
	 black and minima white.
         \label{FIG2}}
\end{figure}

	The underestimation of column density at $A_V > 1$ as derived
from the 60 and 100 \micron\ emission maps comes from an overestimate
of dust temperature along lines of sight that pass through various
environments with different temperatures.  The emission is then
dominated by the warmer regions, while the column density is dominated
by the cooler and denser regions.  The derived color temperature is
thus more representative of the warmer regions, and the derived
optical depth is therefore too low.  This effect can be seen clearly
in the leftmost panel of Figure \ref{FIG1}.  Because $X$ has been
scaled to minimize the difference between the emission and absorption
derived column densities, the derived column densities of points below
$A_V = 1$ are overestimated.  This problem is much less significant
when the emission maps are in the Rayleigh-Jeans portion of dust
emission spectrum \citep{Bethell04}, which explains why the two panels
on the right side of Figure \ref{FIG1} show a much tighter correlation
between the derived column density and the true column density than
the two panels on the left.  Similarly, the longer wavelength emission
maps, shown in Figure \ref{FIG2}, trace the true column density much
better than the shorter wavelength maps.

\subsection{Variable Dust Properties}
\label{VARIABLE}
	In order to isolate the effect of the assumption of isothermal
dust along each line of sight through a molecular cloud on the
emission-derived column density, we have kept the dust properties in
our model invariant.  However, recent observations have shown that the
overall submillimeter emissivity of dust grains varies within a
molecular cloud \citep{Stepnik03} and that the emissivity spectral
index of dust within a molecular cloud also varies \citep{Dupac03}.
The scatter between emission-derived column density and the true
column density shown in Figure \ref{FIG1} arises solely from the
assumption of isothermal dust, and does not include the effects of
variable dust emission properties.  When the column density is derived
from 60 and 100 \micron\ observations, the effect of variable grain
properties may not be important, because the assumption of isothermal
dust makes deriving the column density from emission maps impractical
above $A_V > 1$ while dust properties are not seen to vary
significantly below $A_V < 2$ \citep{Stepnik03}.  When the emission
maps are longer wavelength (e.g. from DIRBE, PRONAOS or SCUBA), the
assumption of isothermal dust is less problematic (as shown in Figure
\ref{FIG1}) and the variable dust properties may become the dominant
source of error in the emission-derived column density.
	
\subsection{Emission from Very Small Grains}
\label{VSG}
	We have shown in \citet{Schnee05} that the 60 \micron\
emission of molecular clouds comes predominantly from transiently
heated very small dust grains (VSG's).  In order to derive the dust
temperature, we have removed the VSG contribution to the 60 \micron\
IRAS flux.  Because we have no way to determine the VSG contribution
on a pixel-by-pixel basis, we assume that the percent emission from
VSG's at 60 \micron\ is constant everywhere in Perseus and Ophiuchus.
We then determine the 60 \micron\ scale factor such that the median
temperature in our maps is equal to the median temperature derived by
SFD98 for the same area.  The SFD98 temperature maps were derived
using 100 and 240 \micron\ data and are therefore unbiased by the
emission from the VSG's.  This procedure is identical to that
performed in \citet{Schnee05}.  The 60 \micron\ flux from our model
was not subjected to such a scaling factor because no VSG contribution
to the flux was included.

	Though scaling the 60 \micron\ flux by a constant
multiplicative factor can give the proper {\it average} temperature,
each line of sight through our observed molecular clouds should
ideally be treated with its own scaling factor.  The variable VSG
contribution to the 60 \micron\ flux can't be removed on a
pixel-by-pixel basis because the SFD98 resolution is too coarse, which
introduces errors in the derivation of dust temperature and column
density.  Applying a constant scale factor to the 60 \micron\ flux
does not correct for this error, nor does it introduce additional
scatter in Figure \ref{FIG3}.  The derived value of $X$ does depend
on this scale factor.

	If high resolution data at 100 \micron\ and longer wavelengths
were available for a molecular cloud, then one could use those maps to
avoid having to subtract the VSG contribution to the observed
emission.  Unfortunately, the dust temperature and column maps
produced by SFD98 using existing 100 and 240 \micron\ data are of much
lower resolution ($\sim$1 degree) than the IRAS maps ($\sim$5').  Even
nearby molecular clouds require the higher resolution provided by
IRAS, so we compare our model molecular cloud with the observations of
Perseus and Ophiuchus using IRAS emission maps.

\subsection{Comparison with Observations}
\label{OBSERVATIONS}
	In real molecular clouds the best estimate of column density
often comes from the measured NIR colors of background stars and is
calculated from the reddening done by the intervening dust.  Column
density maps made this way have a fundamental uncertainty because the
background stars have a spread in their intrinsic NIR colors, and
because the stars included in each pixel of the column density map
sample slightly different lines of sight through the molecular cloud.
The level of uncertainty is approximately 0.2 mag $A_V$, and it is
independent of column density \citep{Ridge05}.

\begin{figure}
\plotone{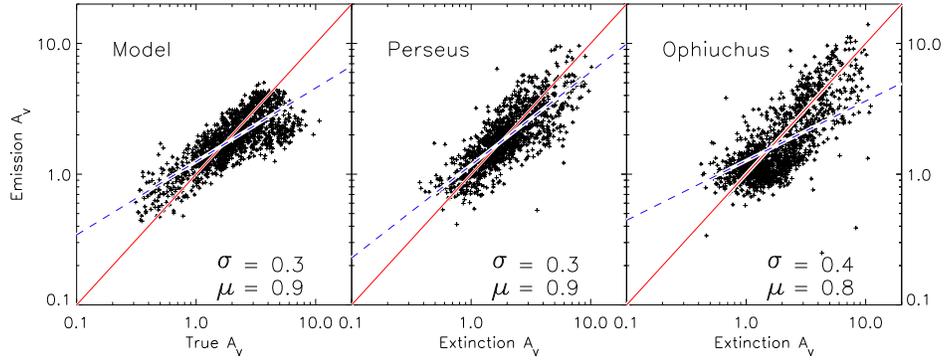}
\caption{The column density derived from 60 and 100 \micron\ 
	 emission maps (assuming $\beta = 2$) plotted against the 
	 ``true'' column density.  The model's true column density has 
	 been altered by gaussian ``noise'' with a 1$\sigma$ of 0.2 
         mag $A_V$ to match the noise in observational estimates of 
	 extinction.  The middle and right panels come from 
	 \citet{Schnee05}, where the extinction is measured
         from NIR colors of backgroud stars and the 60 and 100 \micron\
	 flux from the IRAS survey are used to estimate the column 
	 density from dust emission.  The blue (dashed) line shows the best 
	 fit between the emission-derived column density and the ``true'' 
	 column density.  The slope of the model fit is 0.53 which is 
	 intermediate between the Perseus and Ophiuchus slopes (0.66
	 and 0.46, respectively).  The values of $\mu$ and $\sigma$ refer
         to the center and width of the gaussian fit to the distribution
         of the ratio ($A_{V_{emission}}/A_{V_{true,extinction}}$).
	 \label{FIG3}}
\end{figure}

	In order to compare our simulation to observations, we haved
added normally-distributed noise (with 1$\sigma = 0.2$ mag) to the
simulated column density.  Assuming $\beta = 2$, we then derived the
value of $X$ that best matches the emission-derived column density
(from the 60 and 100 \micron\ maps) to the new ``true'' column
density.  The result is shown in Figure \ref{FIG3}.  

	The left panel of Figure \ref{FIG3} appears similar to the
Perseus and Ophiuchus extinction scatter plots, produced as described
in \citet{Schnee05}, shown here in the center and right panels.  The
emission-based column densities in Perseus and Ophiuchus are derived
from IRAS 60 and 100 \micron\ maps and the extinction is derived from
from NIR color excesses \citep{Ridge05}.  The column derived from FIR
emission maps is overestimated at low extinctions and underestimated
at high extinctions in both this paper and in observations.  The $1
\sigma$ width of the gaussian fit to the ratio of emission-derived
column density and absorption-derived column density are 0.3 in our
simulation, 0.3 in Perseus and 0.4 in Ophiuchus, as shown in Figure
\ref{FIG3}.

	We are able to reproduce the scatter observed in Perseus
between the column density derived from 60 and 100 \micron\ emission
maps and the extinction in our simulation of a turbulent molecular
cloud.  The dust in our simulation does not have variable emission
properties, nor is there a population of transiently heated VSG's.
This suggests that any variations in dust properties within Perseus do
not dominate the uncertainty in the emission-derived column density,
nor does the emission from the VSG's.  Instead, we find that in
Perseus the scatter is primarily explained by the assumption of
isothermal dust along each line of sight.  In contrast, the Ophiuchus
molecular cloud has a larger emission/absorption derived scatter than
can be explained solely by the errors introduced by the assumption of
isothermal dust along each line of sight.

	We have shown in \citet{Schnee05} that in order to produce
column density maps of nearby molecular clouds from IRAS emission
maps, it is necessary to determine both the fraction of 60 \micron\
emission that comes from transiently heated dust grains and the
conversion factor ($X$) from 100 \micron\ optical depth to visual
extinction, because both of these quantities vary from place to place.
These parameters can be determined by the temperature maps of SFD98
and 2MASS/NICER extinction maps.

\section{Summary}

	We have shown that column density estimates from longer
wavelength thermal emission maps are less biased and have less
uncertainty than estimates made from shorter wavelength maps.  In
particular, maps of column density derived from 60 and 100 \micron\
emission maps or 100 and 240 \micron\ emission maps systematically
overestimate the dust temperature at high extinctions ($>$1 mag $A_V$
and 4 mag $A_V$, respectively) and underestimate the column density.

	We have also shown that the observed uncertainty in column
density derived from IRAS emission maps of Perseus can be fully
accounted for by errors introduced by the assumption of isothermal
dust along each line of sight through the cloud and uncertainty in the
extinction-derived column density.  We find no evidence of variable
dust emission properties or variations in the fraction of emission at
60 \micron\ from transiently heated VSG's creating significant scatter
in column density estimates in Perseus.  The observed scatter in the
emission/absorption derived column density in Ophiuchus can be
largely, but not fully explained by the assumption of isothermal dust
along each line of sight, suggesting that the VSG emission and/or dust
properties vary within the Ophiuchus molecular cloud.

\acknowledgments 

	T. Bethell would like to thank Fabian Heitsch for making
available his simulations of MHD turbulence.  We would like to thank
Ellen Zweibel for her numerous insights and helpful suggestions.  The
60/100 \micron\ and NIR extinction maps of Perseus and Ophiuchus
analyzed in this paper were created for the COMPLETE Survey of
Star-Forming Regions \citep{Goodman04,Ridge05}.  Scott Schnee is
supported under a NSF Graduate Research Fellowship.  Tom Bethell is
partially supported by NSF AST 03-28821 and NASA ATP 04 0114 to
U. Wisconsin.  This work has been partially supported by the National
Science Foundation through award AST-0407172.

%\bibliographystyle{apj}
%\bibliography{refs}

\clearpage

\clearpage

\clearpage

\end{document}